\documentclass[12pt]{article}

\title{Non-linear way to Supersymmetry and N-extended SUSY
\footnote{Invited talk, will be published in Proceed. of SUSY-30, Supl. of NP B}
}

\author{V. Akulov
\footnote{Physics Department, Baruch College of the City University of New
York \newline
New York, NY 10010 , USA\newline 
E-mail: akulov@gursey.baruch.cuny.edu}}

\begin{document}

\maketitle

\begin{abstract}
In this report I give the short historical review
some of the first steps that were done to the invention of SUSY
in Kharkov team headed by D.Volkov.

This paper is dedicated to the memory of
Prof. Yu. Gol'fand, whose ideas of SUSY inspired
the most active developments in High Energy Physics over
thirty years.

\end{abstract}

\begin{flushright}
{\it ''Geometry of space is associated with mathematical group''}

Felix Klein , ''Erlagen Program'' 1872
\end{flushright}

This year, the science community celebrates the 30th anniversary of SUSY. I
have been asked to give a short historical introduction to the first steps
in this direction that were done by our group headed by Prof. Dmitry Volkov.

Rochester's High Energy Conference was held in Kiev (Ukraine, Soviet Union) in
1970. Prof. Yu. Golfand announced 2 reports for this conference, but the
Org. Committee gave him time only for one. Prof. Yu. Gol'fand preferred to
discuss the problem of vacuum in QED, because the problem of Superalgebra
Poincar\'e, which was obtained already, seemed to him very complicated for a
first discussion. But the abstract of this report, with Superalgebra Poincar\'e,
was published in the Rotaprint edition of the Proceeding~\cite{/1/}.

It was the first publication about the superalgebra Poincar\'e. Unfortunately,
in the final version of Proceedings did not appear this abstracts.

But the first attempts of the introduction of superalgebras in physics was
given in paper by G.Stavraki \cite{/2/} in 1966 and H.Miyazava \cite{/3/}
in 1968.

The mathematical background for supersymmetry was constructed in 1970.
Felix Berezin(Moscow) and Gregory Kats (Kiev) \cite{/4/}
published the paper about the groups with a commuting and
anticommuting parameters in the Russian journal
''Mathematicheskiy Zbornik'' -- that is how the supergroups appeared in
mathematics.

Before that Prof. D. Volkov investigated the connection between spin and
statistics and rediscovered (after M.Green,1953) \cite{/5/}
the parastatistics in 1959. Later he considered fermionic
Regge trajectory. The success of the application of Goldstone's Theorem
\cite{/6/}  to $\pi$-meson physics stimulated the desire for the
generalization of this theorem to fermionic case. Another hint was linked
with the incorrect Heisenberg's idea about the neutrino as a Goldstone
particle connected with broken discrete symmetry -- P-parity.

I was a post-graduated student at that time and Prof. D. Volkov proposed this
theme for my Ph.D. Thesis in 1971.
Prof. D. Volkov headed the Research Laboratory in Theoretical Physics
Division, which was headed by academician A. Akhieser,
in the Ukrainian Physics and Technology Institute (UPhTI)
in Kharkov. Volkov's laboratory included also the post-graduated
students -- V. Tkach, V. Soroka, L. Gendenshtein,
A. Zheltukhin, V. Gershun, A. Pashnev and later D. Sorokin and
I. Bandos, A. Gumenchuk, A. Nurmagombetov.

The main direction of research activity was connected with the application
of Group theory to the Particle Physics.
E. Cartan's book ''Geometry of Lie Groups and symmetric spaces'' was a main
textbook for us during period.
Only the one who had finished studing of that textbook, could participate in
further activity.
D.Volkov understood that we needed much greatermathematical background to
achieve this goal. As he said: ''I'm afraid to rediscover what is already
dicovered''.

For using Klein \& Cartan's approach, we were needed the new sort of
group -- the group with
anticommuting and commuting parameters.
We participated in the seminars of Mathematical Physics, headed by
Prof. V. Marchenko in the Mathematics
Department of Kharkov State University. D. V. Volkov
discussed with V. Marchenko our problem and one of the participant,
V.Golodets, told us that he has read the paper
of F. Berezin and G.~Kats dealing with new kind of group.

We first attempted the use of the exponential representation for the Poincar\'e
Supergroup, but this approach was more complicated.

However the last part of the paper by Berezin-Kats contained an example of
the matrix realization of supergroup on
the graded Pauli matrices SL$\left( 1|1\right) =\sigma_{0},\sigma_{+},
\sigma_{-}$ with the generalized commutator -- the
supercommutator $\left[ \sigma_{\pm}, \sigma_0\right] =0$,
$\left\{\sigma_{+}, \sigma_{-}\right\} =\sigma_0$

Note,that a superalgebra contained the unit matrix in contrary to the
standard Lee algebra and satisfied generalized the
Jakobi's identity. Then we constructed a matrix representation for a
supergroup $3\times 3$ using graded $\lambda$-matrices by Gell-Man:

$$
\lambda ==\left| 
\begin{array}{ccc}
\lambda _{11} & \lambda _{12} & \lambda _{13} \\ 
\lambda _{21} & \lambda _{22} & \lambda _{23} \\ 
\lambda _{31} & \lambda _{32} & \lambda _{33} 
\end{array}
\right| 
$$
where only $\lambda_{13}, \lambda_{23}, \lambda_{31}, \lambda_{32}$
have a grassmannian parity. That was an examples of graded 
$SU\left( 2\right|1)$.

Then, using the well-known representation for the Poincar\'e
group like an upper triangular matrices 
\begin{equation}
\begin{array}{l}
P=
\left(
\begin{array}{cc}
\begin{array}{cc}
1 & 0 \\
0 & 1
\end{array}
& iT \\
0 &
\begin{array}{cc}
1 & 0 \\
0 & 1
\end{array}
\end{array}
\right) \times 
\left(
\begin{array}{cc}
L & 0 \\
0 & L^{\left( +-1\right) }
\end{array}
\right)
\end{array}
\end{equation}
where each of the block is two by two matrix, and $L=L\left(
l\right)$ is a matrix of the group $SL\left( 2,C\right)$, which
corresponds to the Lorentz group, and $T=T\left( x\right)$ is a
Hermitian matrix corresponding to the group of translations.

We constructed an extended Poincar\'e supergroup, insert to the
center of matrices new blocks

\begin{equation}
\begin{array}{l}
    G = KH =
    \left(
    \begin{array}{ccc}
        1 & \Xi & 1/2\Xi \Xi ^{+}+iT \\
        0 & 1 & \Xi ^{+} \\
        0 & 0 & 1
    \end{array}
    \right)
    \times\left(
    \begin{array}{ccc}
        L & 0 & 0 \\
        0 & U & 0 \\
        0 & 0 & L^{+-1}
    \end{array}
    \right) =\\
    \\
    \phantom{G= KH }
    = \left(
    \begin{array}{ccc}
        L & \Xi U & \left( 1/2\Xi \Xi ^{+}+iT\right) L^{+-1} \\
        0 & U & \Xi ^{+}L^{+-1} \\
        0 & 0 & L^{+-1}
    \end{array}
    \right)
\end{array}
\end{equation}

In the matrices (2), the block $U=U\left(u\right)$ corresponds to the
representation of a unitary group $N\times N$ unitary matrices
with parameters $u$. In the simplest case, $U$ is the one-dimensional
identity matrix. The block $\Xi=\Xi \left( \xi \right)$ is determined by
two indexes: one of them carried by spinor index, corresponding to the group 
$SL\left( 2,C\right)$ and another carried by
unitary index. This block is an arbitrary rectangular matrix with two rows
and $N$ columns. The block $\Xi^{+}\left(\xi\right)$ is
the Hermits conjugate of the block $\Xi\left(\xi\right)$. It is assumed
that the matrices $\Xi$ and $\Xi^{+}$ are linear with respect to the group
spinor parameters $\xi$ and $\xi^{*}$, so that these anticommute with each
other and commute with another parameters.

In the definitions (1) and (2) we have written down representations of the
matrices as products of two factors, each of them determines a certain
group. The principal advantage gained by representing the group as product is
that when one acts on the left
on the products (1) and (2) with matrices of the group the parameters in the
left-hand factors are transformed solely through
themselves and correspond to certain homogeneous spaces.For the Poincar\'e
group (1), the homogeneous space
defined in this manner is ordinary 4-dimensional Minkovsky's space-time.

For the Poincar\'e supergroup (2), the homogeneous space or corresponding
coset contains ordinary space-time as well as additional anticommuting
spinor degrees of freedom -- this space was later called ''superspace'' by
A. Salam and J. Strathdee.

A transformation of the group parameters corresponding to the matrix product
(2)

\begin{equation}
G\left(g^{^{\prime \prime}}\left(g^{^{\prime }},g\right)\right) =
G\left(g^{^{\prime}}\right) G\left(g\right)
\end{equation}
expressed in terms of the parameters in the definition of the
individual blocks, had the form

\begin{equation}
L(l^{^{\prime \prime }}) = L(l^{\prime })L(l)
\end{equation}

\begin{equation}
U(u^{\prime \prime}) = U(u^{\prime})U(u)
\end{equation}

\begin{equation}
\Xi(\xi ^{\prime \prime}) = L(l^{\prime})\Xi(\xi)U^{-1}(u^{\prime
})+\Xi \left( \xi ^{\prime }\right)
\end{equation}

\begin{equation}
\begin{array}{l}
T(x^{\prime \prime }) = L(l^{\prime })T(x)L^{+-1}(l^{\prime }) +
T(x^{\prime }) +
\\
\phantom{T(x^{\prime \prime }) = } 1/2i\left[\Xi(\xi^{\prime})U(u^{\prime })\Xi (\xi
)L^{+-1}(l^{\prime }) - \right.
\\
\phantom{T(x^{\prime \prime }) = asdf}
\left. L(l^{\prime })\Xi \left( \xi \right)
U^{-1}(u^{\prime })\Xi ^{+}(\xi ^{\prime })\right]
\end{array}
\end{equation}

The transformation (4) and (5) correspond to the ordinary transformations
of the Lorentz group and the Unitary group.

The structure of the first term in the transformation (6) is due to the
circumstances already noted that the matrix $\Xi$
has one spinor and one unitary index. The second term corresponds to
translations in the spinor space. In the transformation
(7), the first two terms describe a transformation of translations in the
Poincar\'e group. The last term in (7) establishes
the relationship between translations in the ordinary space and spinor space.

Note that the transformations (6) and (7) do not contain the parameters 
$l^{\prime\prime}, u^{\prime\prime}, l, u$. This means that the variables $x$
and $\xi$
do indeed form a homogeneous space under left shifts in formula (3).In order
to distinguish group parameters from the coordinates of the homogeneous
space, we shall henceforth denote the latter by variables $x$ for
translations in the Poincar\'e
group and $\theta$ for spinor translations.

Expanding the matrices $\Xi$ and $T$ with respect to a complete system of
matrices, we obtain:

\begin{equation}
\Xi\left(\theta\right) = Q_\alpha^k\theta_k^\alpha ,
\end{equation}

\begin{equation}
\Xi^{+}\left(\theta^{*}\right) = Q_{k\alpha *}\theta^{k\alpha *},
\end{equation}

\begin{equation}
T(x) = \widetilde{\sigma}_\mu x^\mu ,
\end{equation}
where $\left(Q_\alpha^k\right)_b^a = \delta_\alpha^a\delta_b^k$,
and $\widetilde{\sigma}_\mu$ are the relativistic Pauli matrices.

We defined a transformation of the coordinates of the homogeneous space
under transformations corresponding to
the parameters $\xi$. Replacing $\xi, \xi^{\prime}, \xi^{\prime\prime},
x, x^{\prime\prime}$ by $\theta, \xi, \theta^{\prime }, x, x^{\prime}$ we
obtained

\begin{equation}
\theta _\alpha ^{i\prime }=\theta _\alpha ^i+\xi _\alpha ^i,
\end{equation}

\begin{equation}
x^{\mu\prime} = x+1/2i\left(\xi^{*}\sigma^\mu\theta -
\theta^{*}\sigma^\mu\xi\right)
\end{equation}

Under transformations of the Poincar\'e group, the $x^\mu$
transform as ordinary coordinates. The transformation of these quantities
under the Lorentz group and the unitary group is determined by their
indices. Note that the spinors $\theta$ transform under all transformations
of $G$ only among themselves.

The matrices $Q_\alpha^i, Q_{i\alpha^{*}}^{+}$ and $P^\mu = 
\widetilde{\sigma}^\mu$ together with the generators of the Lorentz
group and the unitary group form the complete set of the generators of $G$
in this representation. The commutation
relations for these generators can be readily found from the definitions and
have the form

\begin{equation}
\left\{Q_\alpha^i, Q_{k\beta^{*}}^{+}\right\} =
1/2\delta_k^i\sigma_{\alpha\beta ^{*}}^\mu\widetilde{\sigma}_\mu.
\end{equation}

\begin{equation}
\left\{ Q_\alpha ^i,Q_\beta ^k\right\} =
\left\{ Q_{i\alpha^{*}}^{+},Q_{k\beta ^{*}}^{+}\right\} =0.
\end{equation}

\begin{equation}
\left[\widetilde{\sigma}_\mu, Q_\alpha^i\right] = 
\left[\widetilde{\sigma}_\mu, Q_{k\beta^{*}}^{+}\right] = 
\left[\widetilde{\sigma}_\mu, \widetilde{\sigma}_\mu\right] = 0.
\end{equation}

The commutation relations of the operators $Q, Q^{+}$ and $P_\mu =
\widetilde{\sigma}_\mu$ with the generators $L_{\mu\nu}$ of the Lorentz group and
$I_{ik}$ of the unitary group are uniquely determined by their
Lorentz and unitary indexes. At the same time,
operator $\widetilde{\sigma}_\mu$, which is associated with translations
in the Poincar\'e group, satisfies the commutation relations for
the energy-momentum operator. The commutation relations for the operators $Q$
and $Q^{+}$ are determined by
anticommutators, since the corresponding group parameters are anticommuting.

Thus, basing on the extended Poincar\'e supergroup, we constructed a
nontrivial unification of
the space-time symmetries like the Lorentz group or the Poincar\'e group with
the internal symmetries $U\left(n\right)$.
This way we bypassed the ''no-go'' theorem of Coleman-Mandula (The
restriction to Lie group has no a priori grounds).

The above coset $\widetilde{P/H}$ included the grassmannian spinor
coordinates with unitary index
along with the conventional 4-dimensional Minkovski space and was later
called ''Extended Superspace''
by A.Salam and J.Strathdee.

Our next step was the construction of the action integral invariant against
such supergroup.
We introduced the Cartan-Maurer differential 1-form, which are the
coefficients of the supergroup
generators in the expression:
$G^{-1}dG = H^{-1}K^{-1}dK$,
$H+H^{-1}dH = \omega^\mu P_\mu +\omega_\alpha^iQ_i^\alpha +
\omega^{i\alpha *}Q_{i\alpha *}+...$,
where the ellipses correspond to omitted terms with generators unitary group
and Lorentz group.

Expanding the product
$$
K^{-1}dK=
\left(
\begin{array}{ccc}
0 & d\Xi & 1/2\left( d\Xi\Xi^{*} - \Xi d\Xi^{*}\right) + idT \\ 
0 & 0 & d\Xi ^{*} \\ 
0 & 0 & 0 
\end{array}
\right)
$$
with respect to the group generators,we obtained the following expressions
for the Cartan-Maurer
forms: 
\newline
$$
\omega_\alpha^i = d\theta_\alpha^i,
$$ 
$$
\omega^{i\alpha ^{*}} = 
d\theta^{i\alpha ^{*}},
$$ 
$$
\omega^\mu = dx^\mu - 1/2i\left(\theta^{*}\sigma^\mu d\theta -
d\theta^{*}\sigma^\mu\theta\right)
$$

It is readily seen by direct calculations that under the transformations of
super Poincar\'e group this
forms are indeed invariant.

To construct an invariant action integral it is sufficient to consider
combinations of this forms
that are invariant under transformations of the Lorentz group and the
unitary group can be
represented under the condition that $\theta$ is a function of $x$ and 
$\omega \left( d\right) = \omega_\mu \left( \theta, \partial \theta /\partial
x\right) dx^\mu$ in the form:
$$S=\int L\left( \theta ,\partial \theta /\partial x\right) d^4x$$
The presence of the volume element $d^4x$ in the expression for action
imposes important restrictions
on the structure of the accepted combinations of the forms $\omega$.

To restrict the number of possible invariants, we used an additional
requirement that the degree of
the derivatives $\partial \theta /\partial x$ in the action be minimal,
which corresponds to allowing in the $S$-matrix
for only the lowest powers of the momenta of the Goldstone fermions. Among
the invariant outer products
we had the unique product,which contained only the differential forms 
$\omega^\mu$:

$$
S = \int \omega^\mu \wedge \omega^\nu \wedge \omega^\rho \wedge 
\omega^\lambda \epsilon_{\mu \nu \rho \lambda } = 
\int d^4x\det \left| \omega_\nu ^\mu \right| 
$$
that would describe a spontaneous broken supersymmetry. It had the form of
the Born-Infeld action
in the absence of the gauge $F$-field -- as was noted by R. Kallosh in 1997, from
the modern perspective
one can say that it was one of the first version of $D-3$
branes \cite{/7/}.

During the summer of 1972 we finished this work and sent the short version
to JETP Lett.
and Phys.Lett.B. \cite{/8/}. The latter was a great problem for our Institute, because
all papers intended for publication abroad
had to be cleared in Moscow. This would normally take 3 month or more. Only
after the positive decision
from Moscow could we send our paper abroad.

In the autumn of 1972 we attended the International seminar on the 
''$\mu-e$ problem'' in Moscow. Prof. D. Volkov
wanted to attract an attention to new kind of Goldstone particle and to
announced on the possible universal
neutrino interaction (Of course, we did not assume that neutrino is
realistic Goldstone particle, because we used
the $U(1)$ subgroup, but not $SU(2)$). Prof. E. Fradkin invited him to give a
two hour talk at the Theoretical Division of FIAN.
At that time Prof. V. Ogievetsky (who was very close to SUSY idea -- he tried to
consider the Rarita-Schwinger's field as the gauge
field) told us about the paper by Yu. Golfand - E. Likhtman in JETP Lett.,
that contained a similar algebra. Yu. Golfand (who worked in FIAN at that time)
was absent during Volkov's report,
neither did E.Likhtman attend it.

We returned back to Kharkov and read the Golfand-Likhtman's paper, that
contained the super Poincar\'e algebra and the action of an Abelian gauge
model with linear realized supersymmetry. We added the reference about their
paper and send our detailed report to the Russian journal ''Theoreticheskaya
and Mathematicheskaya Fizika''
and published in Kiev preprint \cite{/9/}.

Further development had to connect with the generalization to the local
extended Poincar\'e supergroup.
We announced that the local version would contain also the Rarita-Schwinger
field with spin 3/2 like superpartner of graviton.
I had to finish my Ph.D.thesis at this time.
D.Volkov and V.Soroka constructed the local version, using the supermatrix
approach (2).
Gauge fields was introduced and first version of supergravity appeared in
\cite{/10/}.
But the breakthrough in SUSY was began following the paper by J.Wess and
B.Zumino \cite{/11/}.

In conclusion I would like to thank the Org. Committee and special to Misha
Shifman for the invitation and the support.

\end{document}